\documentstyle[sprocl]{article}
\input{psfig}
\def\Journal#1#2#3#4{{#1} {\bf #2}, #3 (#4)}

\def\PRD{{\em Phys. Rev.} D}

\def\be{\begin{equation}}
\def\ee{\end{equation}}
\def\bea{\begin{eqnarray}}
\def\eea{\end{eqnarray}}

\begin{document}
\title{TIME-VARIATIONS OF SOLAR NEUTRINO SIGNALS}
\author{ P.I. Krastev }
\address{
Institute for Advanced Study, Olden Lane, Princeton,\\ NJ 08540, USA}
\maketitle\abstracts{
The most popular solutions of the solar neutrino problem which assume
massive neutrinos predict varying with time signals in the solar
neutrino detectors. The variations represent a distinctive feature of
these solutions and do not depend on the theoretically predicted total
flux of neutrinos from the Sun. The time scale of the variations
ranges between an hour (MSW) and several years (neutrino magnetic
moment). The amplitude of the variations depends on the neutrino
oscillation parameters. Future detectors with high event rates will be
able to test these predictions at a new level of precision. The
prospects for testing neutrino physics solutions by these detectors
will be discussed and the limits on the neutrino oscillation
parameters that one should be able to obtain will be presented.}
\section{Introduction}
Solar neutrino experiments are entering the stage of precision
measurements. SuperKamiokande~\cite{superk} and SNO~\cite{sno} are
going to measure not only the total boron neutrino flux, which alone
would be very important, but also the spectrum of recoil electrons and
the possible time-variations of the signals. Perhaps the single most
important measurement will be the measurement of $CC/NC$
ratio~\cite{chen} in the SNO detector, which might rule out all ``new
physics'' solutions, should the double ratio,
$(CC/NC)_{exp}/(CC/NC)_{th}$, of the experimentally measured to the
theoretically predicted ratios turn out to be close to
one.\cite{howwell}

The new level of experimental precision requires precise calculations
of the expected signals in the future detectors and a careful analysis
of the possible implications of all conceivable experimental
results. It is well known that astrophysical solutions do not provide
a solution of all the solar neutrino problems,\cite{SNP1} the number
of which, instead of diminishing, seems to increase with every new
solar neutrino experiment. At present at least two of the three
qualitatively different experiments (chlorine, water and gallium) have
to be wrong in order for an astrophysical solution to become
possible.\cite{SNP2}  Even fitting the experiments with solar neutrino
fluxes unconstrained by the complications of solar and nuclear
physics, which is taken care of in ``state of the art'' solar model
calculations, but subject only to the luminosity constraint doesn't
solve the problem.\cite{SNP3} The minimal $\chi^2$ is in an unphysical
region with negative beryllium neutrino flux. On the other hand,
neutrino oscillations provide an excellent fit to the data without
drastically changing solar models or arbitrarily manipulating the
experimental results. Still, a solar model independent signature in
the experimentally measured signals, which would be the ultimate proof
of lepton number violation, has not been seen yet. Such signatures
would be spectral distortions, ratios of different kinds of solar
neutrino event rates, preferably in one and the the same detector,
such as CC/NC, CC/(electron scattering), etc., and time-variations
beyond that due to the 7 \% energy independent effect from the
eccentricity of the earth's orbit. Measurements of time-variations
require real-time detectors with sufficiently high event rates and the
ability to determine the energy of the incident solar neutrinos. An
additional advantage would be the possibility to isolate signals
coming from solar neutrinos with a line spectrum, such as $^7{\rm Be}$
and/or $pep$ neutrinos.

\section{Variations due to vacuum oscillations}
  Variations in this case are due to the eccentricity of the Earth's
orbit.\cite{vacosc} The survival probability is a function which
depends strongly on the distance between source and detector. The
time-variations are particularly strong for neutrinos with line
spectrum, but are large enough to be observed in detectors like SNO
and SuperKamiokande.  Signals in these detectors are predicted to vary
by up to 15 \% between minimum and maximum.\cite{vac1} It is important
to use a flux independent normalization of the experimental results,
e.g. to divide the average event rate during a month (or a week) by
the average event rate during a whole year.\cite{vac2} This will make
possible the study of variations independently of the predicted model
neutrino flux. The variations are quite sensitive to changes in the
detector thresholds, a feature which can be used as an additional
check of their nature. Analysis of data obtained during the same time
intervals by different detectors, e.g. SNO and SuperKamiokande, can
provide further information, thus enhancing the discovery potential of
these detectors.

\section{Variations due to MSW transitions}
Day-night and seasonal variations have been identified a long time ago
as particularly revealing features of the MSW
effect.\cite{{smi},{baltz}} The electron neutrinos in the sun are
supposedly converted into muon or tau neutrinos inside the sun. During
day time these neutrinos are detected on Earth without further
modification. At night however, the neutrinos can undergo an
additional transition, this time from muon or tau back to the original
electron neutrinos, thus enhancing the night-time event rates. This
spectacular feature of a solar neutrino signal can in principle be the
final, still missing, proof of MSW conversions in the sun. Note that
the Kamiokande collaboration has already ruled out a relatively large
portion of the $\Delta m^2$ $-$ $\sin^22\theta$ plane, the one where
this effect is particularly strong, by not seeing any sign of
variability of their signal.\cite{kam}  SuperKamiokande is unlikely to
do much better even with the superb statistics, because of its
sensitivity to the $\nu_\mu$-electron scattering, which reduces the
amplitude of the day-night variations. On the other hand, SNO has a
better chance to probe almost the whole large mixing allowed region
and even some portion of the small mixing-angle allowed region. In
Fig.1 the asymmetry, (night-day)/(night+day), of the signals within
one year between night-time and day-time event rates is plotted in the
$\Delta m^2$ $-$ $\sin^22\theta$ plane for the two detectors. Further
improvement of the sensitivity of these detectors to the day-night
effect can be achieved by analyzing data during shorter time intervals
around the winter solstice when the neutrinos pass for a short time
through the core of the earth.\cite{{baltzW},{losecco}}

\begin{figure} 
\hbox{\psfig{figure=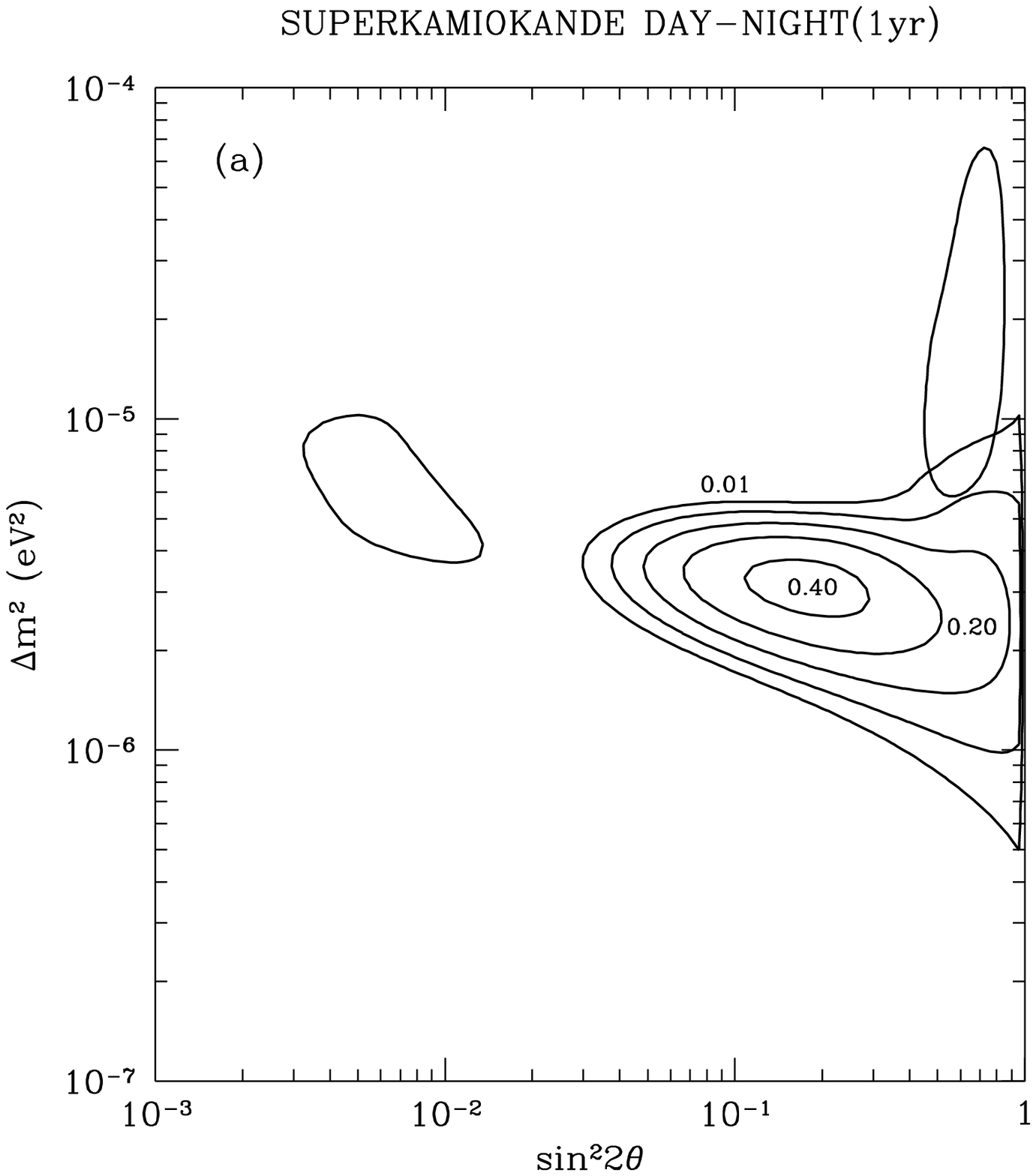,height=2.5in}
\hfill\psfig{figure=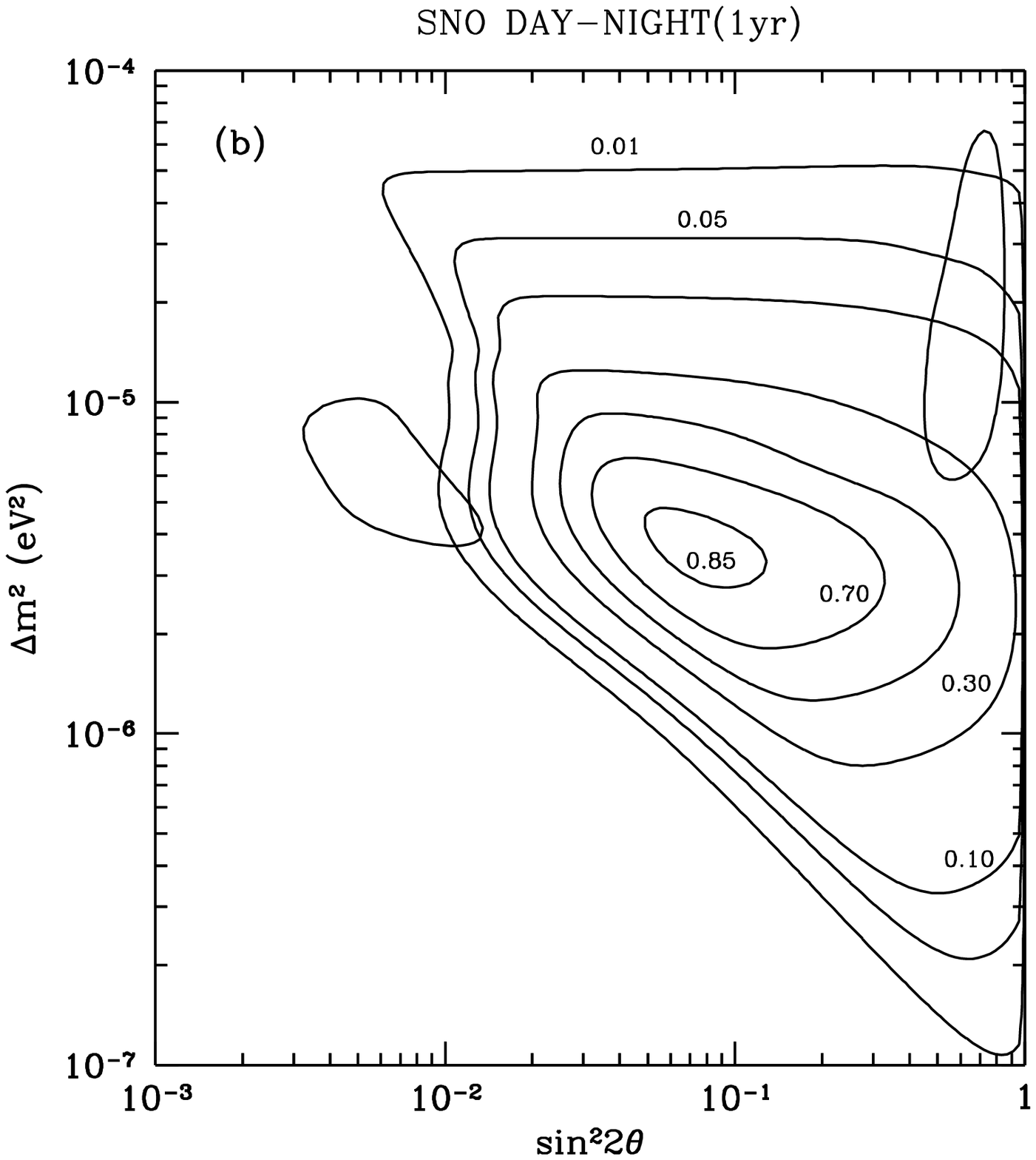,height=2.5in}} 
\caption{
Asymmetry (day-night)/(day+night) in the SuperKamiokande (a) and SNO
(b) detectors.  \label{fig:SKSNO}}
\end{figure}

\vskip -0.5cm
\section{Conclusions}
\vskip -0.2cm
Variations of solar neutrino signals in future detectors might provide
decisive evidence for or against neutrino oscillations.

\vskip -1cm
\section*{Acknowledgments} 
\vskip -0.2cm
This work has been partially supported by funds from the Institute for
Advanced Study.

\end{document}